\def\year{2019}\relax
\documentclass[letterpaper]{article} 
\usepackage{aaai19}  
\usepackage{times}  
\usepackage{helvet}  
\usepackage{courier}  
\usepackage{url}  
\usepackage{graphicx}  
\usepackage[ruled,linesnumbered]{algorithm2e}
\usepackage{amssymb}
\usepackage{amsfonts,amsmath}
\usepackage{mathrsfs}
\usepackage{comment}
\usepackage{subcaption}
\usepackage{color}
\begin{document}
\title{Enhancing Robustness of Deep Neural Networks Against Adversarial Malware Samples: Principles, Framework, and AICS'2019 Challenge}
\author{Deqiang Li \textsuperscript{1,4}, Qianmu Li \textsuperscript{1}, Yanfang Ye \textsuperscript{2}, Shouhuai Xu \textsuperscript{3} \\~\\
\textsuperscript{1} School of Computer Science and Engineering, Nanjing University of Science and Technology
\\
\textsuperscript{2} Department of Computer Science and Electrical Engineering, West Virginia University
\\
\textsuperscript{3} Department of Computer Science, University of Texas at San Antonio
\\
\textsuperscript{4} School of Computing and Information Sciences, Florida International University
\\
lideqiang@njust.edu.cn, qianmu@njust.edu.cn, yanfang.ye@mail.wvu.edu, shouhuai.xu@utsa.edu\\
}
\maketitle
\begin{abstract}
Malware continues to be a major cyber threat, despite the tremendous effort that has been made to combat them. The number of malware in the wild steadily increases over time, meaning that we must resort to automated defense techniques. This naturally calls for machine learning based malware detection. However, machine learning is known to be vulnerable to adversarial evasion attacks that manipulate a small number of features to make classifiers wrongly recognize a malware sample as a benign one. The state-of-the-art is that there are no effective countermeasures against these attacks. Inspired by the AICS'2019 Challenge, we systematize a number of principles for enhancing the robustness of neural networks against adversarial malware evasion attacks. Some of these principles have been scattered in the literature, but others are proposed in this paper for the first time. Under the guidance of these principles, we propose a framework and an accompanying training algorithm, which are then applied to the AICS'2019 challenge. Our experimental results have been submitted to the challenge organizer for evaluation.
\end{abstract}

\section{Introduction}
Malware remains a big threat to cyber security despite communities' tremendous countermeasure efforts. For example, Symantec \cite{symantec:Online} reports seeing 355,419,881 new malware variants in year 2015, 357,019,453 in year 2016, and 669,974,865 in year 2017.
Worse yet, there is an increasing number of malware variants that attempted to undermine anti-virus tools and indeed evaded many malware detection systems \cite{cisco:Online}.

In order to cope with the increasingly severe situation, we have to resort to machine learning techniques for automating the detection of malware in the wild \cite{DBLP:journals/csur/YeLAI17}. 
However, machine learning based techniques are vulnerable to adversarial evasion attacks, by which an adaptive attacker perturbs or manipulates malware samples into adversarial samples that would be detected as benign rather than malicious (see, for example, \cite{Biggio:Evasion,al2018adversarial,DBLP:conf/ijcai/HouYSA18,DBLP:conf/eisic/ChenYB17}).\footnote{The term ``adversarial example'' is often used in the literature. We instead propose using the term ``adversarial sample'' because it is arguably more natural in the context of malware detection, for which we already get used to terms like ``benign sample'' and ``malicious sample''. Corresponding to these two kinds of samples, an adversarial sample is a malicious sample that would be misclassified as a benign one.} The state-of-the-art is that there are many attacks, but the problem of effective defense is largely open. This is indeed the context in which the AICS'2019 malware classification challenge is proposed.

\subsection{Our Contributions}
In this paper, we make the following contributions. First, we propose, to the best of our knowledge, the first systematic framework that aims to enhance the robustness of malware classifiers against adversarial evasion attacks. The framework is designed under the guidance of a set of principles, some of which are known but scattered in the literature (e.g., using an ensemble of classifiers), but others are explicitly proposed for the first time, such as the following. We propose using the capability of the optimal white-box attack to bound the capability of any $\ell_p~(p \geq 1)$ norm based gray-box attack from above, and propose using semantics-preserving representation learning for malware classification. Both the principles and framework should be seen as a starting point and systematically refined in the future. 

Second, we apply the framework to address the AICS 2019 malware classification challenge. Among the 3,133 testing samples in 5 classes, our classifier predicts 2,245 samples in class `0', 461 samples in class `1', 162 samples in class `2', 176 samples in `3', and 89 samples in class `4'. Since we do not know the ground truth of the testing set, which has yet to be announced by the challenge organizer, we cannot tell the effectiveness of the framework at the time for writing the present paper. Instead, we have submitted our classification result to the challenge organizer.

\subsection{Related Work}
Since the present paper focuses on defense against adversarial malware classification, we review existing studies in this topic by emphasizing two complementary approaches: {\em input prepossessing} and {\em adversarial training}.

\subsubsection{Input Prepossessing} 
Input prepossessing transforms the input to a different representation with the aim to reduce the degree of perturbation to the original input.
For example, Random Feature Nullification (RFN) randomly nullifies features in the training and testing phases \cite{wang_2017}; hash transformation leverages a locality-preservation property to reduce the degree of perturbation to the original input~\cite{li2018hashtran}; DroidEye \cite{YeFOSINT-SI-2018} quantizes binary feature representation via count featurization.

Our framework uses binarization to reduce the degree of perturbation, which is inspired by the idea of feature squeezing in the context of image processing~\cite{xu2017feature}. This effectively reduces the perturbation space because there are now only two kinds of perturbations: flipping from `1' to `0' or flipping `0' to `1'. This means that we effectively consider the number of perturbations but not the `scale' of perturbation (i.e., reducing the sensitivity of classifiers to small degrees of perturbation).

\subsubsection{Adversarial Training}
Adversarial training augments the training data with adversarial samples to improve the robustness of classifiers. This idea has been independently proposed in different application settings, including \cite{goodfellow6572explaining,DBLP:journals/corr/KurakinGB16a} and 
\cite{xu2014evasion,DBLP:conf/codaspy/XuZXY13}. In particular, it has been proposed to consider adversarial training with the optimal attack, which {in a sense corresponds to the worst-case scenario and therefore} could lead to classifiers that are robust against the non-optimal attacks \cite{al2018adversarial}. The challenge is of course to find the optimal attack.
In our framework, we use this approach to regularize our model and seek the optimal attack via the gradient descent method.

\section{Review on Adversarial Evasion Attacks against Malware Classification} \label{attack}

\subsection{Basic Idea}
Consider a classifier $f:\mathcal{X} \rightarrow {\mathcal{Y}}$ that takes an unperturbed malware instance ${\bf x}\in{\mathcal{X}}$ as input and correctly outputs its label $y \in {\mathcal{Y}}$. Given $\bf{x}$ that is to be classified, the adversarial evasion attack problem is to manipulate or perturb $\bf{x}$ to an adversarial malware sample ${\bf{x'}}$ such that the malicious functionality of ${\bf x}$ is preserved while satisfying:
\begin{align}
&f({\mathbf x'}) \neq f({\mathbf x}), \label{n-ta} \\ 
\text{s.t.}~~~~&~{\mathbf x}_{\rm lb} \leq {\mathbf x'} \leq {\mathbf x}_{\rm ub} \label{constraint} \\
&~||{\bf x'} - {\bf x}|| \leq \epsilon \label{constraint2}
\end{align}
where ${\mathbf x}_{\rm lb}$ (${\mathbf x}_{\rm ub}$) is the element-wise lower (upper) bound of feature vectors, the box-constraint Eq.\eqref{constraint} means that the manipulation or perturbation cannot violate the constraints imposed by the feature definitions and 
${\bf u} \leq {\bf v}$ means that each element of ${\bf u}$ is no greater than the corresponding element in ${\bf v}$, and $||\cdot||$ refers to a metrics of interest (e.g., $\ell_{p}$ norm for some $p\in \{0, 2, \infty\}$)
and Eq.\eqref{constraint2} says that 
perturbations may be bounded by a given $\epsilon$ in the norm.
The perturbation vector is $\delta_{\bf x}={\bf x'}-{\bf x}$. 

In the present paper, we focus on classifiers $f$ that are learned as neural networks, ${\bf F}:\mathcal{X}\rightarrow\mathbb{R}^o$, which output (softmax) the probability mass function over $o$ classes or labels.
Since the constraint given by Eq.\eqref{n-ta} is hard to formulate,  
researchers proposed considering two scenarios:
in the case of {\it non-targeted} attacks, maximize the cost of classifying the ${\bf x'}$ as $y$ 
\begin{equation}
    \max \limits_{\mathbf x'}~L({\bf F}({\mathbf x'}), y) \label{non_tar};
\end{equation}
in the case of {\it targeted} attacks, minimize $L({\bf F}({\mathbf x'}), y_t)$, where $y_t$ ($y_t \neq y$) is a target label given by the attacker.

\subsection{Threat Model} 
As elaborated below, a threat model against malware classifiers is specified by {\em what the attacker knows}, {\em what the attacker can do}, and {\em how the attacker wages the attack}.

\subsubsection{What the attacker knows} 
There are three kinds of models from this perspective.
A {\it black-box} attacker knows nothing about classifier $f$ except what is implied by $f$'s responses to the attacker's queries.
A {\it white-box} attacker knows all kinds of information about $f$, including its model parameters.
A {\it gray-box} attacker knows an amount of information about $f$ that resides in between the preceding two extremes. For example, the attacker may know the training set or feature definitions.

\subsubsection{What the attacker can do}
In evasion attacks, the attacker only can manipulate the testing data, while obeying some constraints.
One constrain is to preserve the malicious functionality of a malware.
Although the attacker can manipulate a malware sample by inserting, deleting, and replacing features \cite{dang2017evading,anderson2017evading}, a simplifying assumption is to consider insertion only (i.e., flipping a feature value from `0' to `1'
\cite{Biggio:Evasion,rndic_laskov,grosse2017adversarial,wang_2017,DBLP:journals/corr/RosenbergSRE17,Chen:2017:SES,al2018adversarial}.
The other constraint is to maintain the relation between features. 
Using the ACIS'2019 malware classification challenge as an example, 
we note that $n$-gram (uni-gram, bi-gram, and tri-gram) features reflect sequences of Windows system API calls. This means that when the attacker inserts an API call into a malware sample, several features related to this API call will need to be changed according to the definition of $n$-gram features.

\subsubsection{How the attacker wages the attack}
Researchers generate adversarial malware samples using various machine learning techniques such as genetic algorithms, reinforcement learning, generative networks, feed-forward neural networks, decision trees, and Support Vector Machine (SVM)~\cite{316904628,anderson2017evading,Hu2017,Biggio:Evasion,xu2014evasion,rndic_laskov,carliniW16a}. In order to generate adversarial malware samples effectively and efficiently, attacks often leverage the gradients with respect to inputs of neural network~\cite{goodfellow6572explaining,papernot_2016}.

\section{Framework}\label{method}

\subsection{Guiding Principles}

The design of the framework is guided by a number of principles. These principles are geared towards neural network classifiers, which are chosen as our focus because deep learning techniques are increasingly employed in malware defense, but their vulnerability to adversarial evasion attacks has yet to be tackled~\cite{raff2017malware}.

\subsubsection{Principle 1: Knowing the enemy}

This principle says that we should strive to extract useful information about the data as much as we can. This kind of information will offer insights into designing countermeasures. For example, we can ask questions of the following kinds:
\begin{itemize}
    \item Is the training set imbalanced? If the training set is not balanced, various methods need to be considered for alleviating the imbalance issue. For example, the widely-used oversampling is to expand the samples of the ``minority'' classes via random and repetitive sampling \cite{buda2018systematic}. 
    \item Are there sufficiently many samples? This issue is important especially when there are a large number of features and when neural networks are considered. One widely-used method is data augmentation, which generates new samples by making slight modifications on original samples \cite{lemley2017smart}. 
    \item Are there simple indicators of adversarial samples? If there are simple indicators of adversarial samples, we can possibly design tailored classifiers for them.
\end{itemize}

\subsubsection{Principle 2: Bridging the gap between countermeasures against gray-box attacks and countermeasures against white-box attacks}
In gray-box attacks, the attacker knows some information about the feature set and therefore can train a surrogate 
classifier $\hat{f}:\mathcal{X}  \rightarrow \mathcal{Y}$ from a training set (where the realization of $\hat{f}$ is a neural network $\hat {\bf F}$) and leverage the transferability from $\hat{f}$ to $f$ to generate adversarial samples. 
{Consider an input ${\bf x}$ for which a gray-box attacker generates perturbations using
\begin{equation}
    \hat \delta_{\bf x} \in \max \limits_{||\hat \delta_{\bf x}||\leq\epsilon}L({\bf \hat F}({\bf x + \hat \delta_x}), y), \nonumber
\end{equation}
the change to the loss of $f$ incurred by $\hat \delta_{\bf x}$ is
\begin{align*}
    \left|\Delta L\right| &= \left|L({\bf F}({\bf x} + \hat \delta_{\bf x}), y) - L({\bf F}({\bf x}), y)\right|\\
    &=\left|\int_{0}^{\hat \delta_{\bf x}}{\triangledown L({\bf F}({\bf x}+\delta), y)}d\delta\right|\\
    &=\left|\int_{0}^{1}\triangledown L({\bf F}({\bf x}+t\hat \delta_{\bf x}), y)^\top \hat \delta_{\bf x}dt\right|\\
    &\leq\epsilon\sup \limits_{||\delta||\leq\epsilon}\left\|\triangledown L({\bf F}({\bf x} + \delta), y)\right\|_*,
\end{align*}
where ``$||\cdot||_*$'' means the dual norm of $||\cdot||$. The preceding observation indicates that corresponding to the same (and potentially large) perturbation upper bound $\epsilon$, the loss incurred by gray-box attacks is upper bounded by the loss incurred by white-box attacks. This suggests us to focus on the robustness of classifier $f$ against the optimal white-box attack because it accommodates the worst-case scenario.
It is worth mentioning that our observation, which applies to an arbitrary perturbation upper bound $\epsilon$, enhances an earlier insight that holds for a small perturbation upper bound $\epsilon$ \cite{demontis2018intriguing}.}

\subsubsection{Principle 3: Using ensemble of classifiers rather than a single one (i.e., not putting all eggs in one basket)}

This is suggested by the observation that no single classifier may be effective against all kinds of evasions. Worse yet, we may not know the kinds of evasions the attacker uses to generate adversarial samples. Therefore, we propose using {\it ensemble} learning to enhance the robustness of neural networks based malware classifiers.

An ensemble can be constructed by many methods (e.g., bagging, boosting or using multiple classifiers). The generalization error of an ensemble decreases significantly with the ensemble size when the base classifiers are effective and mutual independent {\cite{doi:10.1080/01621459.1963.10500830}}. For example, {\it random subspace}~\cite{709601_rss} is seemingly particularly suitable for formulating malware classifier ensembles because the dimension of malware feature vectors is often very high, which indicates a high vulnerability of malware classifiers to adversarial samples~\cite{simon2018adversarial}. 

Since the output of a neural network (with softmax) is the probability mass function over the classes in question, the final prediction result is produced according to these probabilities. Formally, an ensemble $f_{en}:\mathcal{X}\rightarrow\mathcal{Y}$ contains a set of neural network classifiers $\{f_i\}_{i=1}^l$, namely $f_{en} = \{f_i:\mathcal{X}\rightarrow\mathcal{Y};~(1\leq i \leq l)\}$. Given a testing sample ${\bf x}$, each classifier $f_i$ defines a conditional probability on predicting $y$ : $P(y|{\bf x}, f_i)$. We treat the base classifier equally, and the voting method is 
\begin{equation}
    P(y|{\bf x})= \frac{1}{l} \sum_{i=1}^{l}P(y|{\bf x}, f_i),
\end{equation}
where $P(y|{\bf x})$ is the probability that the ensemble predicts ${\bf x}$ as  the label $y$.

\subsubsection{Principle 4: Using input transformation to reduce the degree of perturbation caused by evasion manipulation}
In the context of malware classification, input transformation techniques, such as adversarial feature selection~\cite{zhang2016adversarial} and random feature nullification~\cite{wang_2017}, can reduce the degree of perturbation in adversarial samples so as to improve the robustness of classifiers. 
In typical applications, the defender does not know what kinds of evasion manipulations are used by the attacker to generate adversarial samples, including the number of features that are manipulated by the attacker. Therefore, we propose considering a spectrum of evasion manipulations, from manipulating a few features (measured by, for example, the $\ell_0$ norm) to manipulating a large number of features (but the magnitude of these manipulations may be small and therefore measured for example by the $\ell_\infty$ norm). Moreover, we may give higher weights to the transformation techniques that can simultaneously reduce the degrees of perturbations in terms of the $\ell_{\infty}$ norm, $\ell_0$ norm, or $\ell_2$ norm. This suggests us to propose using the {\it binarization} technique: 
When the feature value of the $i$th feature, denoted by $x_i$, is smaller than a threshold $\Theta_i$, we binarize $x_i$ to 0; otherwise, we binarize $x_i$ to 1.
The threshold $\Theta_i$ may be set as the median value of the $i$th feature. This input transformation reduces the perturbation space to two kinds: flipping `0' to `1' and flipping `1' to `0'.

\subsubsection{Principle 5: Using adversarial training to ``inject'' immunity into classifiers}
Adversarial training (also known as proactive training \cite{xu2014evasion}) incorporates some adversarial samples into the training set. Various kinds of {\em heuristic} training strategies have been proposed (see, e.g., \cite{grosse2017adversarial,szegedyZSBEGF13,xu2014evasion,goodfellow6572explaining,kurakin2016adversarial}.
However, these strategies typically deal with some specific evasion methods and therefore are not known to effective against other evasion methods.
Inspired by robust optimization, \cite{madry2017towards} propose solving the saddle point problem so as to improve the robustness of neural networks against a wide range of adversarial samples. This strategy has been adapted to the context of malware classification in \cite{al2018adversarial}.

The preceding discussion suggests us to train neural networks that can accommodate the optimal attack and the unperturbed examples on distribution $\mathcal{D}$:
\begin{equation}
    \min \limits_{\theta}~\mathbb{E}_{({\bf x}, y)\in \mathcal{D}}\left[\max \limits_{\delta_{\bf x}\leq{\epsilon}}L({\bf F}({\bf x} + \delta_{\bf x}), y) + L({\bf F}({\bf x}),y)\right] \label{loss:clf},
\end{equation}
where $L(\cdot, \cdot)$ is the cross-entropy and $\theta$ denotes the parameters of neural network ${\bf F}$ (as the realization of $f_i$). A key issue is how to deal with the optimal attack. \cite{madry2017towards} showed that the projected gradient descent (PGD) can solve the inner maximum problem effectively for calculating the first-order adversarial samples. Inspired by this insight, we leverage gradient descent on the negative cross-entropy to generate adversarial samples. In order to avoid local minima, we can randomly repeat several times for each point and pick the point minimizing the negative cross-entropy as the initial point.

\subsubsection{Principle 6: Using semantics-preserving representations}
Adversarial malware samples must assure that a manipulated sample is still a malware (i.e., preserving the malicious functionality of the original malware sample). This suggests us to strive to learn neural network models that are sensitive to malware semantics, but not the perturbations because the latter must preserve the malicious functionality of the original malware.
Specifically, we propose using {\it denoising autoencoder} to learn semantics-preserving representations because they can make neural network less sensitive to perturbations. A denoising autoencoder $ae=d \circ e$ unifies two components: an encoder $e:\mathcal{X}\rightarrow{\mathcal{H}}$ that maps an input $M({\bf x})$ to a latent representation ${\bf r} \in \mathcal{H}$ and a decoder $d: \mathcal{H}\rightarrow{\mathcal{X}}$ that reconstructs ${\bf x}$ from ${\bf r}$, where the $\mathcal{H}$ is the latent representation space and $M$ refers to some operations applied to ${\bf x}$ (e.g., adding Gaussian noises to ${\bf x}$). \cite{vincent2010stacked} showed that the lower bound of the {\it mutual information} between ${\bf x}$ and ${\bf r}$ is maximized when the reconstruction error is minimized. 
In the case of Gaussian noise $\epsilon \sim \mathcal{N}(0, \sigma^2)$ 
and reconstruction loss
\begin{equation}
    \mathbb{E}_{\epsilon \sim N(0, \sigma^2)}\left\lVert ae({\bf x} + \epsilon) - {\bf x})\right\rVert_2^2,
\end{equation}
\cite{alain2014regularized} showed that the optimal $ae^*({\bf x})$ is
\begin{equation}
    ae^*({\bf x})=\frac{\mathbb{E}_\epsilon \left[p({\bf x} - \epsilon)({\bf x} - \epsilon)\right]}{\mathbb{E}_\epsilon \left[p({\bf x} - \epsilon)\right]} \label{eq:opt-ae},
\end{equation}
where $p(\cdot)$ is the probability density function. Eq.\eqref{eq:opt-ae} says that representations of a well-trained denoising autoencoder are insensitive to ${\bf x}$ because of the weighted average from the neighbourhood of ${\bf x}$, which is reminiscent of the {\em attention} mechanism~\cite{luong2015effective}.

\begin{figure}[!htbp]
	\centering
	\scalebox{0.42}{
	\includegraphics{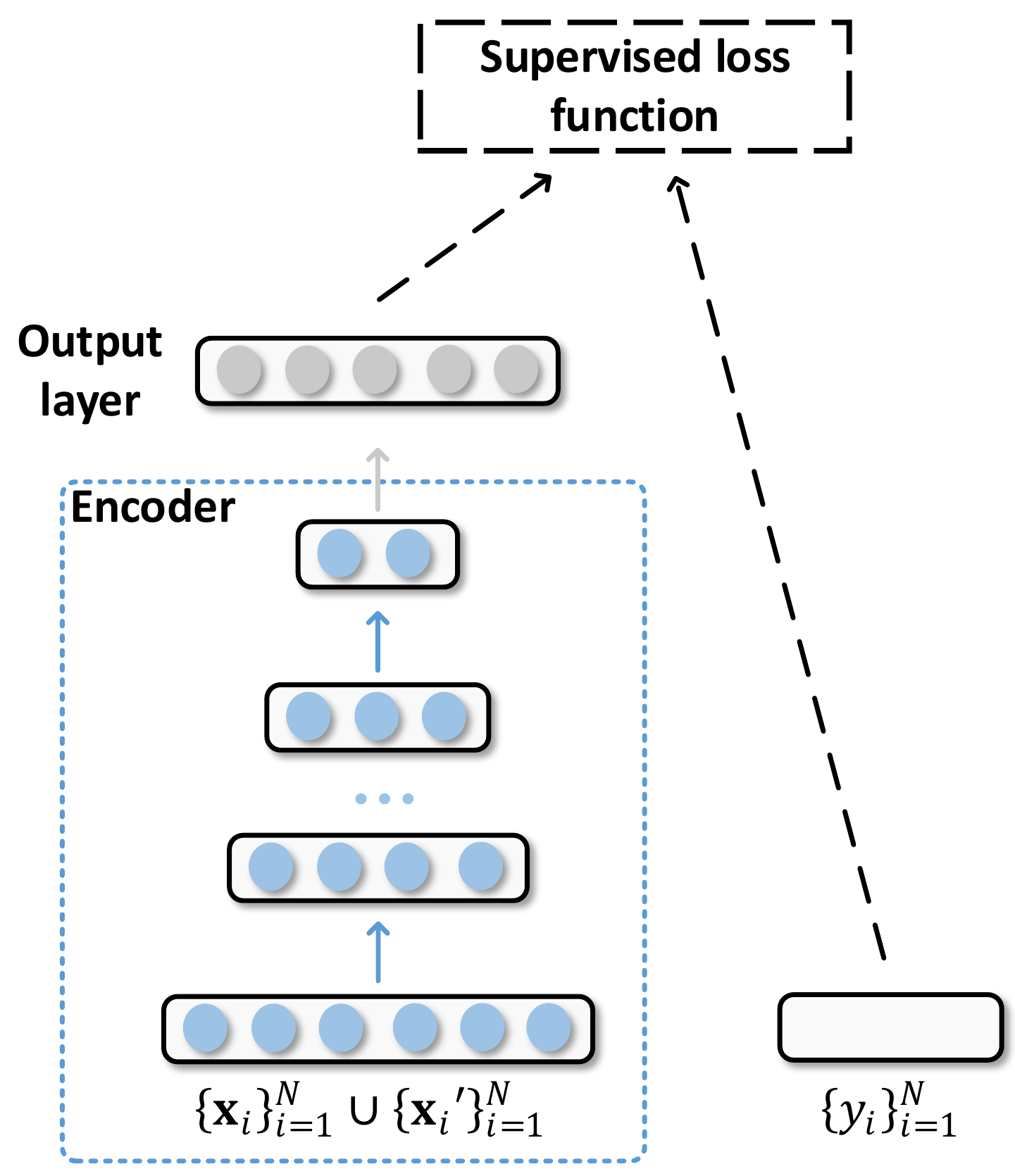}
	}
\caption{Illustration of neural network for classification. Blue dashed box contains the learned encoder, which and the output layer comprise the learned neural network classifier.
The model parameters are tuned to minimize the supervised loss function.}
	\label{fig:clf}
\end{figure}

Figure \ref{fig:clf} depicts how the learned encoder is leveraged for classification, where the encoder structure is the same as described in \cite{vincent2010stacked}. Two examples of noise are:
\begin{itemize}
    \item {\it Salt-and-pepper noise}: A fraction $\alpha$ of the elements of original sample ${\bf x}$ are randomly selected, and then set their values as their respective minimum or maximum (i.e., effectively flipping `1' to `0' or flipping `0' to `1').
    \item {\it Perturbation $\delta_{\bf x}$}: A perturbation $\delta_{\bf x}$ is added to ${\bf x}$ such that classifier $f_{i}$ misclassifies adversarial sample ${\bf x'}={\bf x} + \delta_{\bf x}$.
\end{itemize}
Note that when an input transformation technique mentioned above is used together with a denoising autoencoder, the former should be applied first and the latter is applied to the transformed input.

Given a mini-batch of $N$ training samples ${\bf x}_1, {\bf x}_2, \cdots, {\bf x}_N$, the empirical risk of a denoising autoencoder is
\begin{equation}
    L_{ae}=\frac{1}{N}\sum_{i=1}^{N}\left[\left\lVert ae(M({\bf x}_i))-{\bf x}_i\right\rVert_2^2 + \left\lVert ae({\bf x}'_i)-{\bf x}_i\right\rVert_2^2 \right] \label{loss:dae},
\end{equation}
where $M(\cdot)$ denotes adding salt-and-pepper noise on ${\bf x}$ and ${\bf x}'_i$ is the adversarial sample generated via adversarial training.

\begin{figure*}[!htbp]
	\centering
	\scalebox{0.45}{
	\includegraphics{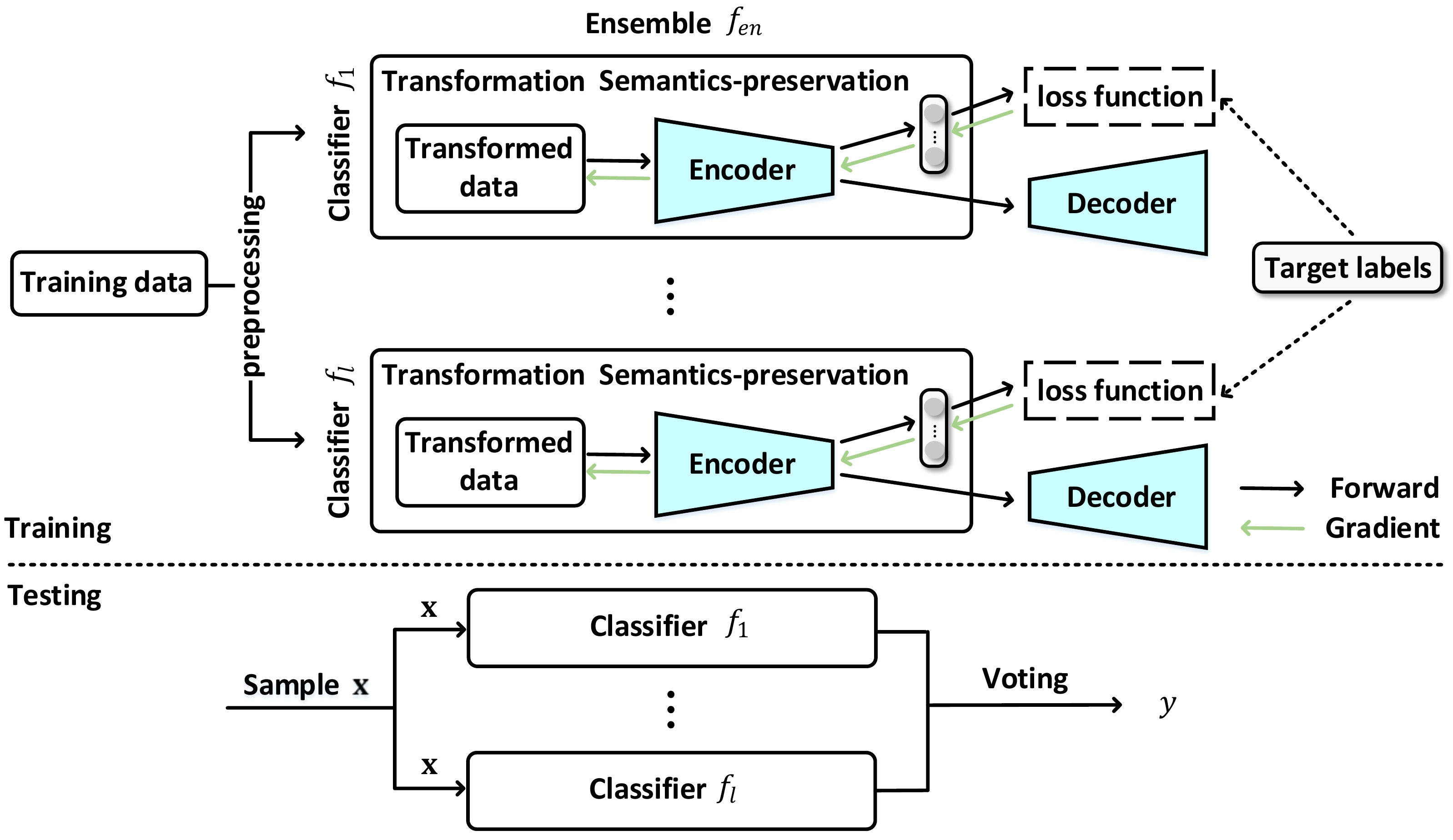}
	}
\caption{Overview of the proposed malware classification framework. In the training phase, an ensemble of $l$ neural network classifiers are trained, with each classifier hardened by three countermeasures (i.e., input transformation, semantics-preserving, and adversarial training using gradient descent on the transformed data). In the testing phase, the label of a sample is determined according to the result of voting by the $l$ classifiers. }
	\label{fig:wf}
\end{figure*}

\subsection{Turning principles into a framework}

The principles discussed above guide us to propose a framework for adversarial malware classification, which is highlighted in Figure \ref{fig:wf} and elaborated below. Specifically, we need to examine whether or not the input has some issues (e.g., imbalanced data) that need to be coped with via an appropriate prepossessing (according to Principle 1). 
We propose using an ensemble $f_{en}$ of classifiers $\{f_i\}_{i=1}^l$ (according to Principle 3), which are trained from random subspace of the original feature space. Each classifier $f_i$ is hardened by three countermeasures: input transformation via binarization (according to Principle 4); adversarial training models on the optimal attacks using gradient descent (green arrows in Figure \ref{fig:wf}, according to Principle 2 and Principle 5); semantics-preservation is achieved via an encoder and a decoder (according to Principle 6). In order to attain adversarial training and at the same time semantics-preservation, we learn classifier $f_i$ via block coordinate descent to optimize different components of the model. 

\begin{algorithm}[!htbp]
\KwIn{Training set $(X, Y)$, maximum training epoch $N_{epoch}$, mini-batch size $N$, and the number of repeat times $K$.}

Cope with issues like imbalanced input;

Select a ratio $\Lambda$ of sub-features to the feature set;

Transform input $X$ to $\overline{X}$ via binarization;

\For{$epoch=$ \rm ${0}$ to $N_{epoch}$}{
Sample a mini-batch $\{{\bf x}_i, y_i\}_{i=1}^N$ from the $(\overline{X}, Y)$;
\BlankLine

\For{$repeat=$ \rm ${0}$ to $K$}{
Apply slight salt-and-pepper noise to $\{{\bf x}_i\}_{i=1}^N$;

Calculate perturbations $\{\delta_{{\bf x}_i}^{repeat}\}_{i=1}^N$ for manipulating sample $\{{\bf x}'_i\}_{i=1}^N$;

Project $\{{\bf x}'_i\}_{i=1}^N$ into the binary space;
}
Select the best perturbation $\delta_{{\bf x}_i}$ from $\delta_{{\bf x}_i}^{repeat}$ where $0 \leq repeat < K$ for ${\bf x}_i~(1 \leq i \leq N)$ so as to minimize the negative cross-entropy;

\BlankLine

Calculate the reconstruction loss via Eq.\eqref{loss:dae};

Backpropagate the loss and update the denoising autoencoder parameters;

\BlankLine

Calculate the adversarial training loss via Eq.\eqref{loss:clf};

Backpropagate the loss and update classifier parameters;
}

\caption{Training classifier $f_i$}
\label{alg:train}
\end{algorithm}

Putting the pieces together, we obtain Algorithm \ref{alg:train} for training individual classifiers. The training procedure consists of the following steps. (i) Given a training set $(X, Y)$, we randomly select a ratio $\Lambda$ of sub-features to the feature set, and then transform $X$ into ${\overline{X}}$ via the binarization technique discussed above. (ii) We sample a mini-batch $\{{\bf x}_i, y_i\}_{i=1}^N$ from $(\overline{X}, Y)$, and calculate the adversarial samples ${\bf x}'_i$ for ${\bf x}_i \in \{{\bf x}_i\}_{i=1}^N$ according to Lines 5-11 in Algorithm \ref{alg:train}). (iii) We pass the $\{M({\bf x}_i)\}_{i=1}^N$ and $\{{\bf x}'_i\}_{i=1}^N$ through the denoising autoencoder to compute the reconstruction loss with respect to the target $\{{\bf x}_i\}_{i=1}^N$ via Eq.\eqref{loss:dae}, and update the parameters of the denoising autoencoder. (iv) We pass the $\{{\bf x}_i\}_{i=1}^N$ and $\{{\bf x}'_i\}_{i=1}^N$ together through the neural networks to compute the classification error with respect to the ground truth label $\{y_i\}_{i=1}^N$ via Eq.\eqref{loss:clf}, and update the parameters of the classifier via backpropagation. Note that Steps (ii)-(iv) are performed in a loop. The output of the training algorithm is a neural network classifier.

\section{Experiment: Applying the Framework to the AICS'2019 Challenge} \label{exp}

\subsection{The AICS'2019 Challenge}

The challenge is in the context of adversarial malware classification (i.e., labeling the class to which a malware sample belongs or {\em multiclass classification}), namely constructing evasion-resistant, machine learning based malware classifiers. 
The dataset, including both the training set and the testing set, consists of 
Windows malware samples (or instances), each of which belongs to exactly one of the following five classes: {\em Virus}, {\em Worm}, {\em Trojan}, {\em Packed malware}, and {\em AdWare}. 

For each sample, the features are collected by the challenge organizer via  dynamic analysis, including the Windows API calls and further processed unigram, bigram, and trigram API calls. The feature names (e.g., API calls) and the class labels are ``obfuscated'' by the challenge organizer as integers, while noting the obfuscation preserves the mapping between the features and the integers representation of them. For example, three API calls are represented by three unique integers, say 101, 102, and 103; then, a trigram API call ``101;102;103'' means a sequence of API calls 101, 102, and 103.  
In total there are 106,428 features.

The testing set consists of adversarial samples and non-adversarial samples (i.e., unperturbed malware samples).
Adversarial samples are generated by a variety of perturbation methods, which are not known to the participating teams. 
However, the ground truth labels of the testing samples are not given to the participating teams.
This means that the participating teams cannot calculate the accuracy of their detectors by themselves. Instead, they need to submit their classification results (i.e., labels on the samples in the testing set) to the challenge organizer, who will calculate the classification accuracy of each participating team.

\subsection{Basic Analysis}

As discussed in Principle 1 of the framework, our basic analysis aims to identify some basic characteristics that should be taken into consideration when adapting Algorithm \ref{alg:train} to this specific case study.


\subsubsection{Is the training set imbalanced?}
\begin{figure}[!htbp]
	\centering
	\scalebox{0.55}{
	\includegraphics[width=0.8\textwidth]{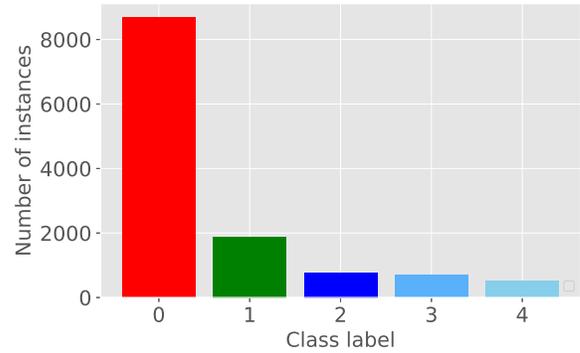}
	}
\caption{Histogram of the training set in malware classes.}
	\label{fig:01}
\end{figure}

The training set consists of 12,536 instances, and the testing set consists of 3,133 instances. The training set contains 8,678 instances in class `0', 1,883 instances in class `1', 771 instances in class `2', 692 instances in class `3', and 512 instances in class `4'. Figure \ref{fig:01} plots the histogram of the instances in the training set according to the given labels of malware class (i.e., the five malware classes that have been obfuscated as integers `0', `1', `2', `3', `4'). We can calculate the maximum ratio between the number of instances in different classes is 16.95, indicating that the training set is highly imbalanced. 

In order to cope with the imbalance in the training set, we use the {\it Oversampling} method to replicate randomly selected samples from a class with a small number of samples. The replication process ends until the number of samples is comparable to that of the largest class (i.e., the class with the largest number of samples), where ``comparable'' is measured by a predefined ratio. In order to see the effect of this ratio, we use a 5-fold cross validation on the training set to investigate the impact of this ratio. The classifier consists of neural networks with two fully-connected layers (each layer having 160 neurons with the ReLU activation function), which are optimized via Adam \cite{kingmaB14} with epochs 50, mini-batch size 128, learning rate 0.001. The model is selected when achieving the best Macro F1 score on the validation set.

\begin{table}[!htbp]
\caption{Accuracy (\%) and Macro F1 score (\%) are reported with a 95\% confidence interval with respect to the ratio parameter (\%), where `---' means learning a classifier using the original training dataset.}
		\centering
		\begin{tabular}{ccc}
			\hline
			 {Ratio (\%)}&{Accuracy (\%)}&{Macro F1 (\%)}\\
			\hline
             ---&93.20$\pm$1.04&85.52$\pm$1.12 \\
            30 &92.86$\pm$0.75&85.47$\pm$1.04 \\
            40 &92.38$\pm$1.00&84.87$\pm$1.07 \\
            50 &92.21$\pm$0.60&84.87$\pm$1.00 \\
            60 &92.48$\pm$1.12&84.62$\pm$1.01 \\
		    \hline
		\end{tabular}
		\label{tab:imbalanced}
\end{table}

Table \ref{tab:imbalanced} shows that the Macro F1 score decreases as the oversampling ratio of minority classes increases. 
In order to make each mini-batch of training samples contain samples from all classes, which would be critical in muticlass classification, our experience suggests us to select the 30\% ratio.

\subsubsection{Are there sufficiently many samples?}
Machine learning, especially deep learning, models need to be trained with a ``large'' number of samples, where ``large'' is relative to the number of features. In the challenge dataset, the training set contains 12,536 samples while noting that there are 106,428 features, which may lead to overfitting. To cope with this, we note that adversarial training is a data augmentation method~\cite{goodfellow6572explaining}, which can be  leveraged to regularize the resulting classifiers.

\subsubsection{Are there simple indicators of adversarial samples?}
In the first testing set published by the challenge organizer, we see negative values for some features. These negative values would indicate that they are adversarial samples. In the revised testing set provided by the challenge organizer, there are no negative feature values, meaning that there are no simple ways to tell whether a sample is adversarial or not. In spite of this, we can speculate the count of perturbed features by comparing the number of nonzero entries corresponding to the samples in the training set to their counterparts in the testing set. It is important because the attack success rate increases with the number of perturbed features.

\begin{figure}[!htbp]
  \begin{subfigure}[b]{0.254\textwidth}
    \includegraphics[width=\textwidth]{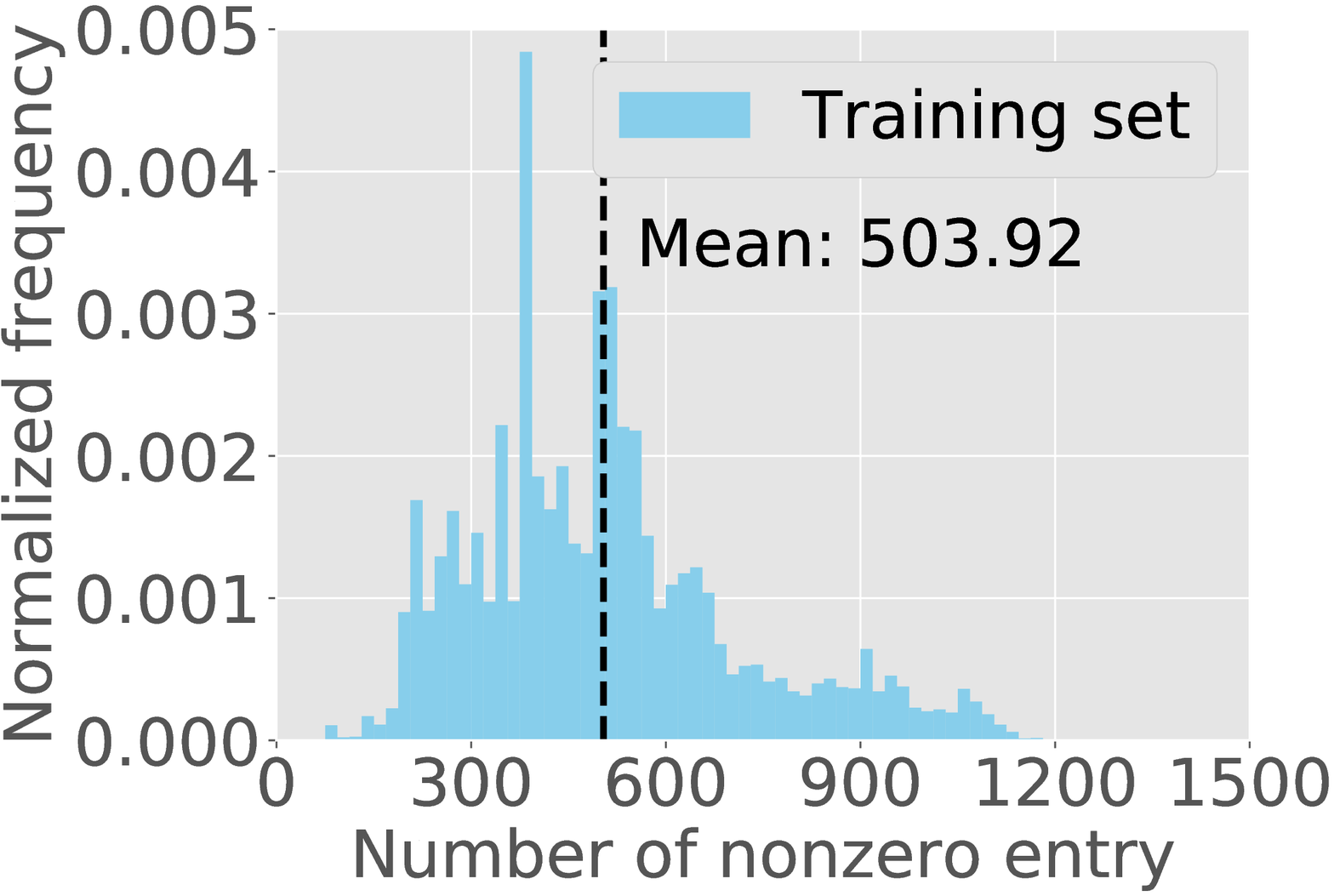}
    \label{fig:train_nonzero}
  \end{subfigure}
  \begin{subfigure}[b]{0.21\textwidth}
    \includegraphics[width=\textwidth]{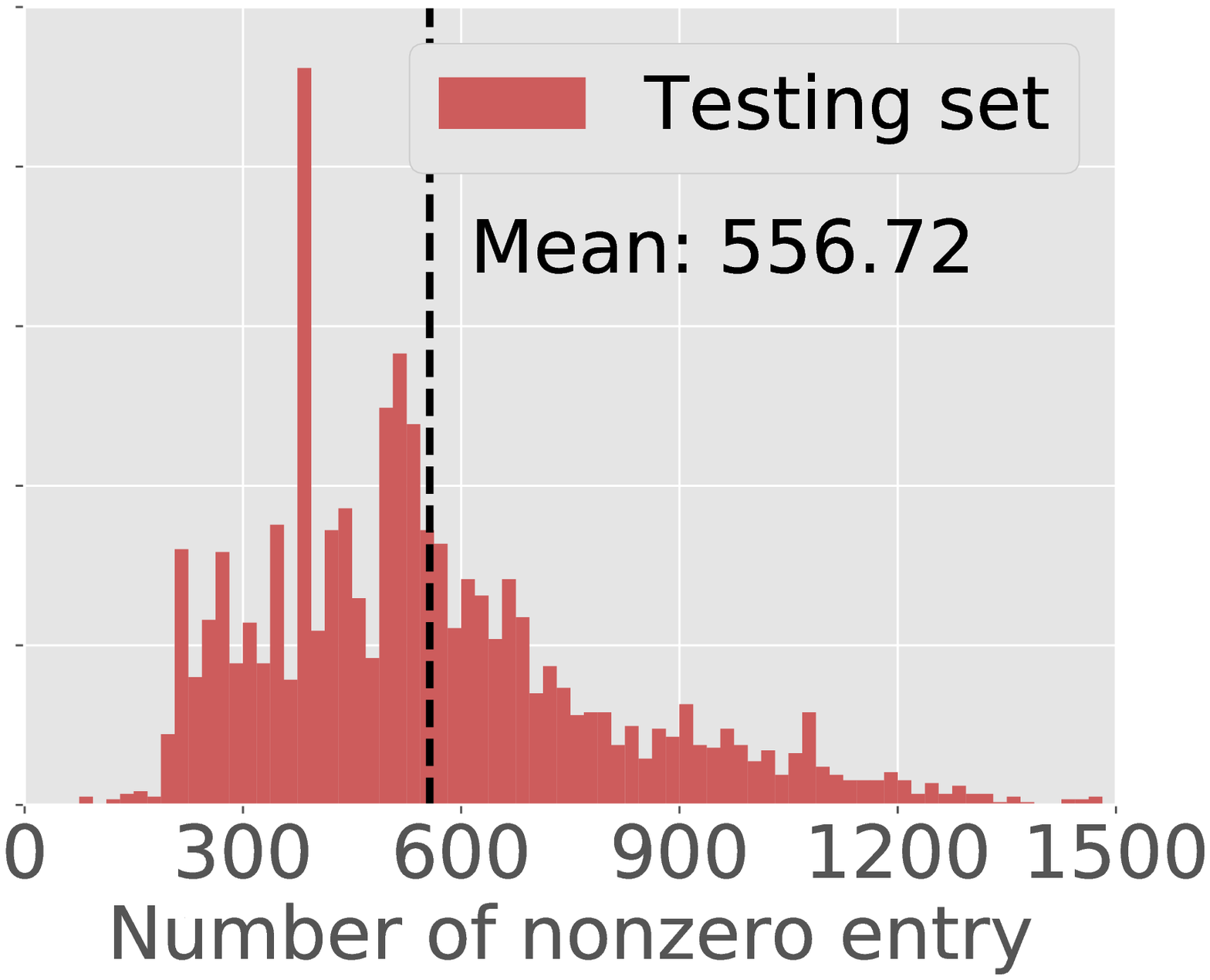}
    \label{fig:test_nonzero}
  \end{subfigure}
\caption{Histogram of the normalized frequency of the number of nonzero entries corresponding to samples in the training set and their counterparts in the testing set. The dashed line represents the mean value.}
  \label{fig:nonzero}
\end{figure}

Figure~\ref{fig:nonzero} shows the normalized frequency of the number of nonzero entries corresponding to the samples in the training set and their counterparts in the testing set. We observe that their normalized frequencies are similar except that some testing samples have more nonzero entries ($>1200$). Their mean values are much smaller than the input dimension ($106,428$), suggesting that the average number of perturbed features may be small.

\subsection{Classification Result}

For adversarial training, the gradient descent with respect to the transformed input iterates 55 times via Adam optimizer~\cite{kingmaB14} with learning rate 0.01. The perturbed input is projected into the range $[0,1]$ and rounded into binary space (i.e., binarization as discussed in the framework). Since we do not have access to the malware samples, we cannot tell whether a feature perturbation preserves the malware functionality or not.
We train 10 neural network based classifiers to formulate an ensemble, including 6 classifiers using the input transformation, adversarial training, and semantics-preservation techniques discussed in the framework, and the other 4 classifiers using the input transformation and adversarial training techniques because some samples may be perturbed without preserving the malicious functionality in the training. The ratio for random subspace method is set as $\Lambda=0.5$. Each classifier has two fully-connected hidden layers (each layer having neurons 160), uses the ELU activation function,
and is optimized by Adam. The classification result has been submitted to AICS 2019 organizer for evaluation. More models and codes are available at \url{https://github.com/deqangss/aics2019_challenge_adv_mal_defense}.

\section{Conclusion} \label{conclusion}

We have systematized six principles for enhancing the robustness of neural network classifiers against adversarial evasion attacks in the setting of malware classification. These principles guided us to design a framework, which leads to a concrete training algorithm. We applied the training algorithm to the AICS'2019 challenge, and submitted the classification result to the challenge organizer for evaluating the effectiveness of our framework.

The problem of adversarial malware detection has not received the due amount of attention. We hope this paper will inspire more research into this important problem. 
Future research problems are abundant, such as: extending and refining the principles against evasion attacks, seeking principles to enhance robustness of adversarial malware detection against poisoning attacks, designing more systematic frameworks and more robust techniques against adversarial malware.

\smallskip

\noindent{\bf Acknowledgments}.
We thank the AICS'2019 challenge organizers for preparing the challenge. DL is supported by the China Scholarship Council under Grant 201706840123.
YY and SX are supported in part by NSF Grant SaTC-1814825. YY is additionally supported in part by NSF Grants CNS-1618629, CNS-1814825 and OAC-1839909, NIJ 2018-75-CX-0032, WV HEPC Grant (HEPC.dsr.18.5), and WVU RSA Grant (R-844). SX is additionally supported in part by ARO Grant W911NF-17-1-0566. 
The views and opinions of the author(s) presented in the paper do not reflect, in any sense, those of the funding agencies.

\bibliographystyle{aaai} 
\bibliography{malware_paper}

\end{document}